\documentclass[onecolumn,draftcls]{IEEEtran}
\pdfoutput=1
\IEEEoverridecommandlockouts
\usepackage{cite}
\usepackage{amsmath,amssymb,amsfonts}
\usepackage{amsthm}
\usepackage{bm}
\usepackage{algorithmic}
\usepackage{graphicx}
\usepackage{textcomp}
\usepackage{xcolor}
\usepackage{subfiles} 
\usepackage{caption} 
\usepackage{enumitem}

\def\BibTeX{{\rm B\kern-.05em{\sc i\kern-.025em b}\kern-.08em
    T\kern-.1667em\lower.7ex\hbox{E}\kern-.125emX}}
    
\newtheorem{theorem}{Theorem}

\newtheorem{definition}{Definition}

\newcommand{\bc}{\mathbf{c}}
\newcommand{\bx}{\mathbf{x}}
\newcommand{\f}{\mathbf{f}}
\newcommand{\bK}{\mathbf{K}}
\newcommand{\bA}{\mathbf{\tilde{A}}}
\newcommand{\bz}{\mathbf{z}}
\newcommand{\bW}{\mathbf{W}}
\newcommand{\bb}{\mathbf{b}}

\newcommand{\bd}{\mathbf}

\newcommand{\bl}{\bm{\ell}}
\newcommand{\btau}{\bm{\tau}}
\newcommand{\blambd}{\bm{\lambda}}
\newcommand{\bmu}{\bm{\mu}}
\newcommand{\bnu}{\bm{\nu}}
\newcommand{\btheta}{\bm{\theta}}

\newcommand{\R}{\mathbb{R}}

\newcommand{\diag}{\textit{diag}}
\DeclareMathOperator{\conv}{convhull}
\newcommand{\either}{\textrm{either}}
\newcommand{\tor}{\textrm{or}}

\newcommand{\gradi}{\nabla_{\bl}g_{\btheta}({\bl}^{i})}
\newcommand{\hgradi}{\nabla_{\bl}g_{\hat{\btheta}}({\bl}^{i})}
\newcommand{\hgrad}{\nabla_{\bl}g_{\btheta}({\bl}^{\textit{new}})}
\newcommand{\grad}{\nabla_{\bl}g_{\btheta}({\bl})}

\newcommand{\tell}{\bl^{\textrm{new}}}

\graphicspath{ {./pictures/} }
\begin{document}

\title{A Convex Neural Network Solver for DCOPF with Generalization Guarantees}
\author{Ling Zhang, Yize Chen and Baosen Zhang
\thanks{The authors are with the Department of Electrical and Computer Engineering at the University of Washington. Emails: \{lzhang18,yizechen,zhangbao\}@uw.edu}
\thanks{The authors are partially supported by NSF grants ECCS-1942326 and ECCS-2023531, and the Washington Clean Energy Institute.}
}
\maketitle

\begin{abstract}

The DC optimal power flow (DCOPF) problem is a fundamental problem in power systems operations and planning. With high penetration of uncertain renewable resources in power systems, DCOPF needs to be solved repeatedly for a large amount of scenarios, which can be computationally challenging. As an alternative to iterative solvers, neural networks are often trained and used to solve DCOPF. These approaches can offer orders of magnitude reduction in computational time, but they cannot guarantee generalization, and small training error does not imply small testing errors. In this work, we propose a novel algorithm for solving DCOPF that guarantees the generalization performance. First, by utilizing the convexity of DCOPF problem, we train an input convex neural network.  Second, we construct the training loss based on KKT optimality conditions. By combining these two techniques, the trained model has provable generalization properties, where small training error implies small testing errors. In experiments, our algorithm improves the optimality ratio of the solutions by a factor of five in comparison to end-to-end models.

\end{abstract}

\section{Introduction}
The optimal power flow (OPF) problem is a fundamental tool used in the planning and operation of power systems~\cite{dommel1968optimal,baldick2006applied,Glover17}. The OPF problem finds the least cost generator outputs that satisfy the power flow equations and other operational constraints. In this paper we specifically consider the DCOPF formulation of the OPF problem, which linearizes the power flow equations~\cite{stott2009dc}. 

The DCOPF problem has been studied extensively for almost half a century and is a workhorse of the power industry. If the generator cost is linear, the DCOPF is a linear program (LP). These LPs can be solved efficiently by a variety of algorithms, which have been implemented in a number of software packages~\cite{zimmerman2010matpower,milano2005open}. Today, a DCOPF problem can be solved quickly for fairly large networks~\cite{stott2009dc}.

Even though solving a single instance of DCOPF problems is easy, computational challenges are arising because of the increase in renewable resources, since they introduce significant uncertainties into generation and load~\cite{tang2017real,mohagheghi2018survey}. To account for these uncertainties, operators often need to repeatedly solve the DCOPF problem for a large number of scenarios~\cite{hauswirth2016projected,zhang2017dynamic,huang2018fast}. If these scenarios are analyzed close to real-time, then using standard solvers can become too inefficient. For example, suppose a single instance of DCOPF can be solved in 1 second. Then solving a thousand instances would require more than 15 minutes, which would likely be too slow. 

Recently, end-to-end neural network architectures have been proposed as surrogates to conventional LP solvers~\cite{pan2019deepopfa,pan2019deepopfb,deka2019learning,roald2019implied,zamzam2019learning,chen2020learning,fioretto2020predicting}. These neural networks treat load as the input and output the generation values. Since they only require simple function evaluations, they offer orders of magnitude speedup compared to iterative algorithms. 
Despite the increase in speed, these machine learning approaches lack provable guarantees on their performances. There are two broad classes of approaches using neural networks for DCOPF. In~\cite{pan2019deepopfa,pan2019deepopfb,zamzam2019learning,fioretto2020predicting}, the neural network is used to directly approximate the functional mapping from load to the optimal generations. In~\cite{deka2019learning,roald2019implied,chen2020learning}, neural networks are used to identify the binding constraints and the solutions are recovered by solving linear systems of equations. Both of these approaches rely on training with a large set of labeled data, then showing the performance of the algorithms through simulations. 

A fundamental barrier in adopting these methods in practice is the need to show that small \emph{training error} implies small \emph{testing errors}. That is, we need to show that the method \emph{generalizes}.  Using machine learning for DCOPF is most useful when operators are faced with unfamiliar conditions, but there are many examples when machine learning precisely fails in these conditions~\cite{cohn1994improving,Papernot16,cobbe2018quantifying}. The hesitancy in using machine learning (especially deep learning) methods also comes from the perception that they rely on ``black-boxes" that are hard to understand~\cite{Womble18,cremer2019optimization}. 

In~\cite{chen2020learning}, a two-step approach for solving DCOPF is proposed.
Firstly, a neural network is used to learn the \emph{value} (i.e., the optimal cost) of the DCOPF, and its gradient with respect to the load is the locational marginal prices (LMPs). Then the binding constraints are identified based on the LMPs. This process is robust in the sense that if the LMPs are learned well, and the binding constraints are correctly identified. However, the question on how well the neural network generalizes on learning the LMPs was not answered. In this paper, we address the generalization question by directly designing the fundamental features of the DCOPF problem into the machine learning algorithm. Specifically, we modify the approach in~\cite{chen2020learning} by 1) constraining the neural network architecture, and 2) building KKT conditions into the training process.

The first technique we use to improve generalization is to constrain the neural network to have an input convex structure, since the cost of the DCOPF is a convex function of the loads. Therefore, we use input convex neural networks (ICNNs) to learn this relationship~\cite{Amos17,Chen19,Chen20voltage}. 
It turns out that the convex structure of the neural network allows it to generalize well.
We show that a small training error for a finite number of samples implies that the error would be small for entire regions of inputs. Intuitively, this means that the gradient of the function (the LMPs) would be roughly correct as long as some points in the input region are sampled during training. This result is in contrast with standard generalization results in the literature, where most are of a statistical nature~\cite{keskar2016large}. Instead, we show that the structure of the neural network is the key, since if convexity is not imposed, we can construct cases with zero training error but arbitrary large testing errors. 

The second technique we use is to add KKT conditions to the training process. Perhaps the most direct way to improve generalization is to increase the number of labeled samples. However, for even moderately large power systems, covering the whole load space with labeled data is intractable due to the curse of dimensionality. We overcome this challenge by using that at optimality, the primal and dual variables satisfy the KKT conditions~\cite{boyd2004convex}. Interpreting the dual variables as the partial derivatives of the value function with respect to different parameters, the KKT conditions can be written as a set of partial differential equations. 
We train the ICNN by minimizing the training loss that is based on this set of partial differential equations. 
This enable us to include a much larger set of inputs without explicitly adding labeled training data.



In summary, our contributions are:
\begin{enumerate}[topsep=-5pt]
    \item We constrain the neural network to have an input convex structure, which allows the model to generalize well. The guarantees on generalization performance are given in Theorem \ref{thm:same_region} and 
    Theorem \ref{thm:unseen_region}.
    These theorems prove our method can generalize to testing data points from spaces unseen in the training process.
    The effectiveness of our method in improving generalization performance is also demonstrated through simulations. In the testing stage, more than $93\%$ of the solutions obtained from our model are optimal, while more than 50\% of the end-to-end solutions are infeasible.
    
    \item We add a term based on violations of KKT conditions to the regression loss, enabling us to use large amounts of unlabeled samples for training and further improving the generalization performance. When we test on a region where no training data is sampled from, the solutions from our algorithm is 97\% optimal, compared to 9\% optimal ratio of end-to-end models.

\end{enumerate}

The challenge in solving DCOPF repeated for different loads is similar in spirit to solving linear systems of equations for changing right-hand-side vectors. Methods like LU factorization have been used to efficiently find these solutions to types of problems in power systems for several decades~\cite{crow2015computational}. In this paper we are trying to solve a parametric optimization problem rather than a parametric linear system~\cite{ji2017operation}.  However, unlike LU factorization, it is hard to guarantee that the machine learning methods would always recover the right answer. The approach in~\cite{venzke2020learning} can bound the worst case errors for a trained neural network with fixed parameters. But these types of ex post analysis is hard to generalize and do not shed light on why a method may perform better or worse. In this paper, we show how designing the structure of the neural network can lead to more robust guarantees. 



This paper is organized as follows. Section~\ref{sec2} provides the DCOPF model and Section~\ref{sec3} reviews the solution method in~\cite{chen2020learning}. Section~\ref{sec4} describes the neural network design and the training loss based on KKT conditions, and Section~\ref{sec5} states and proves the generalization guarantees. Simulations illustrating the results are shown in Section~\ref{sec6}.


\section{DCOPF} \label{sec2}

\subsection{Model}
Consider a power system where the $n$ buses are connected by $m$ edges. For each of the bus, we let $x_i$ denote the output of the generator located at the bus, and let $l_i$ denote the load consumed at the bus. Let $\bx=(x_1,\dots,x_n)$ and $\bl=(l_1,\dots,l_n)$ be the generation and load vectors, respectively. The generation cost vector is denoted as $\bc\in\R^{n}$. We assume that $\bc$ is non-negative and has at least one positive component. We assume the system is connected. For notational simplicity, we assume that all buses have generation and load. Without loss of generality, we assume $x_i$ is bounded by $0$ and $\bar{x}_i$.\footnote{Since the power flow equations are linear, nonzero lower bounds can be shifted to be zero by subtracting a constant from the generation values.} If bus $i$ does not have any generation capability, we set $\bar{x}_i=0$. 

The line flows are related to the bus power injections through a linear relationship. Because of Kirchhoff's laws, not all flows in the $m$ lines are independent. In particular, there are only $n-1$ fundamental flows and the rest of $m-n+1$ flows are linear combinations of the fundamental ones~\cite{zhang2014network,diestel2012graph}.  Let $\f\in\R^{n-1}$ denote these $n-1$ fundamental flows. More details on line flow modeling is given in Appendix~\ref{app:flow}.


The DCOPF problem asks for the least cost generations while satisfying all the loads and flow constraints:
\begin{subequations} \label{P1}
\begin{align} 
J^{\star}(\bl) =\min_{\bx, \f}~& \bc^T \bx\\
\textrm{s.t.~} & \mathbf{0} \leq \bx \leq \bar{\bx} \label{1a}\\
& -\bar{\f} \leq \bK\f \leq \bar{\f} \label{1b}\\
& \bx + \bA\f = \bl , \label{1c}
\end{align}
\end{subequations}
where the matrix $\bK\in\R^{m\times (n-1)}$ maps $\f$ to the flows on all edges, and the matrix $\bA\in\R^{n\times (n-1)}$ is the modified incidence matrix that maps the fundamental flows to the nodal power injections. The value of the optimization problem is denoted as $J^{\star}(\bl)$ and the optimal solution is denoted as $\bx^{\star}(\bl)$.

The DCOPF problem in \eqref{P1} is an LP and is readily solved by most optimization packages such as CPLEX or Gurobi~\cite{zimmerman2010matpower}. These solvers have been optimized to the point that a single LP can be solved in the order of seconds even for large systems. The challenge comes from the fact that repeatedly solving \eqref{P1} for changing loads can be time-consuming, even if the time it takes for a single instance is small. 

\subsection{Example} \label{sec:example}
\begin{figure}[ht]
    \centering
    \includegraphics[scale=0.4]{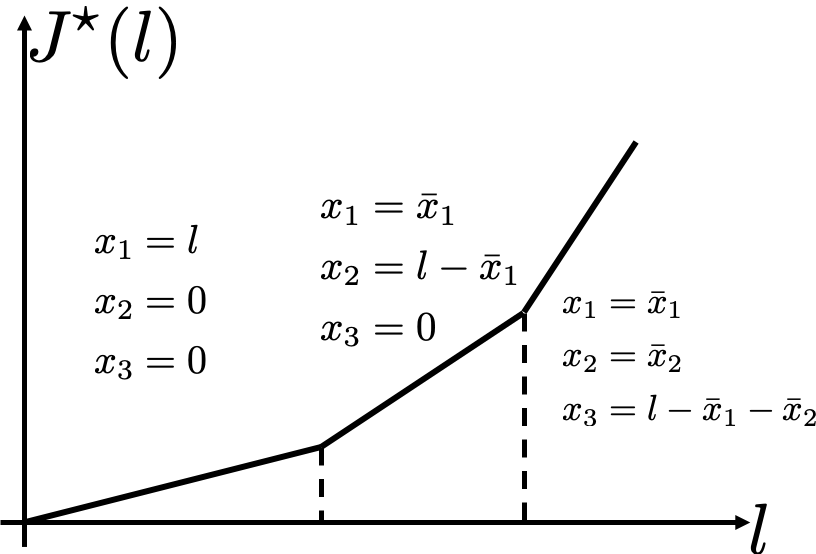}
    \caption{The cost curve of a single bus load with three generators. The curve is piecewise linear, convex and increasing, with each piece corresponding to a different generation profile.}
    \label{fig:single_bus}
\end{figure}

we present a small example that is used to illustrate many of the points in this paper. Consider a single-bus system with three generators (with cost \$1/MW, \$2/MW and \$3/MW, respectively) serving a load. The DCOPF problem becomes 
\begin{subequations} 
\begin{align}
    J^{\star}(l)=\min \; & x_1+2x_2+3x_3 \\
    \mbox{s.t. } & 0 \leq x_i \leq \overline{x}_i, i=1,2,3 \\
    & x_1+x_2+x_3=l. 
\end{align}
\end{subequations}
Figure~\ref{fig:single_bus} plots the cost against the load. The curve is piecewise linear, convex and increasing, with slopes of $1$, $2$ or $3$ depending on the value of the load. Different pieces correspond to different generation profiles as shown in Fig.~\ref{fig:single_bus}.  In an end-to-end approach, the goal is to learn the generations directly. In the next section we will introduce a method that learns the curve $J^{\star}(l)$ and its derivatives. 
 

\section{Learning Active Constraints}\label{sec3}
The goal of using machine learning is to avoid resolving \eqref{P1} every time the load changes. A number of algorithms have been proposed to directly learn the functional mapping from $\bl$ to $\bx^{\star}(\bl)$~\cite{deka2019learning,roald2019implied,chen2020learning}. However, it is difficult to ensure the learned solutions satisfy the constraints in \eqref{1a} to \eqref{1c}. For the example in Fig.~\ref{fig:single_bus}, each of the generations have upper and lower bounds as well as the load balance constraint (sum of generation is equal to the load). In this section we review the algorithm in~\cite{chen2020learning}, where instead of using end-to-end neural networks, the mapping $J^{\star}(\bl)$ is learned. Then the associated dual variables are obtained from the global dependence on the right-hand-side vector in the LP. With the value of the dual variables, we are able to find out a set of active constraints for (\ref{P1}). The exact value of $\bx^{\star}(\bl)$ is then found by solving a system of linear equations.


\subsection{Global dependence on the right-hand side vector}
The global dependence of the optimal cost function $J^{\star}(\bl)$ on the load vector $\bl$ can be found through standard duality theory \cite{boyd2004convex}. The Lagrangian associated with (\ref{P1}) is 
\begin{multline}\label{def:Lagrangian}
L(\bx, \f, \underline{\btau}, \bar{\btau}, \underline{\blambd}, \bar{\blambd}, \bmu) = \bc^T\bx - \underline{\btau}^T\bx + \bar{\btau}^T(\bx - \bar{\bx}) -\\ \underline{\blambd}^T(\bar{\f}+\bK\f) +\bar{\blambd}^T(\bK\f - \bar{\f})
+\bmu^T(\bl-\bx - \bA\f),
\end{multline}
where $\underline{\btau}, \bar{\btau}\in\R^{n}$ are the dual variables associated with generator capacity constraints (\ref{1a}), $\underline{\blambd}, \bar{\blambd}\in\R^{m}$ are the dual variables associated with flow capacity constraints (\ref{1b}), and $\bmu\in\R^{n}$ are the dual variables associated with equality constraints (\ref{1c}). The dual variables $\bmu$ are called the locational marginal prices (LMPs) since they represent the marginal cost of supplying one more unit of power at a bus~\cite{kirschen2018fundamentals}.   

The dual problem of (\ref{P1}) is 
\begin{align}
\max_{\underline{\btau}, \bar{\btau}, \underline{\blambd}, \bar{\blambd}, \bmu}~& \bmu^T\bl - \underline{\blambd}^T\bar{\f} -\bar{\blambd}^T\bar{\f} - \bar{\btau}^T\bar{\bx} \label{P2}\\
\textrm{s.t.}\quad  & \bc - \underline{\btau} + \bar{\btau}-\bmu = \mathbf{0} \tag{\ref{P2}{a}}\label{2a}\\
& - \bK^T\underline{\blambd} + \bK^T\bar{\blambd} - \bA^T\bmu = \mathbf{0}\tag{\ref{P2}{b}}\label{2b}\\
& \underline{\btau}, \bar{\btau}, \underline{\blambd}, \bar{\blambd} \geq \mathbf{0} \tag{\ref{P2}{c}}. \label{2c}
\end{align}
We assume that the load $\bl$ is feasible unless stated otherwise. The key observation from~\cite{chen2020learning} is that from the value of $\bmu$, we can learn the active constraints set for (\ref{P1}). To be specific, the optimal solutions for (\ref{P1}) are associated with the following active/inactive constraints through the value of $\bmu^{\star}$:
\begin{equation}\label{gen-active}
	x_i^{\star} = \left\{
	\begin{array}{ll}
		0, & \textrm{if~}\mu_i^{\star}-c_i<0\\
		\bar{x}_i, & \textrm{if~}\mu_i^{\star}-c_i>0\\
		(0, \bar{x}_i), & \textrm{otherwise},\\
	\end{array}
	\right.
\end{equation}
and
\begin{equation} \label{flow-active}
	f_i^{\star} = \left\{
	\begin{array}{ll}
		\bar{f}_i, & \textrm{if~}\bar{\lambda}_i^{\star}-\underline{\lambda}_i^{\star}>0\\
		-\bar{f}_i, & \textrm{if~}\bar{\lambda}_i-\underline{\lambda}_i^{\star}<0\\
		(-\bar{f}_i, \bar{f}_i), & \textrm{otherwise},\\
	\end{array}
	\right.
\end{equation}
where, given the value of $\bmu^{\star}$, the value of $\bar{\lambda}_i^{\star}-\underline{\lambda}_i^{\star}$ can be determined by solving the following optimization problem:
\begin{align} \label{P3}
	\min_{\bnu}~ & \|\bnu\|_1 \\
	\textrm{s.t.~} & \bK^T \diag(\bar{f}_1,\cdots,\bar{f}_m) \bnu = \bA^T\bmu, \nonumber
\end{align}
where $\diag(\cdot)$ is a diagonal matrix. The value of $\bar{\lambda}_i^{\star}-\underline{\lambda}_i^{\star}$ is related to $\bnu$ by $\bar{\lambda}_i^{\star}-\underline{\lambda}_i^{\star}=\nu_i/\bar{f}_i$~\cite{chen2020learning}.

It turns out the learning problem becomes much simpler from the dual problem. Instead of directly learning the optimal solutions or the active constraints (neither are continuous in the load), it suffices to learn $J^{\star}$, which is a scalar function that is continuous in $\bl$. The multipliers $\mu$ can be then recovered from the following theorem about the global dependence of the optimal cost $J^{\star}(\bl)$ on load $\bl$: 
\begin{theorem}
\label{sub-gradients}
A vector $\bmu^{\star}$ is an optimal solution to the dual problem (\ref{P2}) if and only if it is a (sub)gradient of the optimal cost $J^{\star}(\bl)$ at the point $\bl$, that is, 
\begin{equation}\label{def:dev}
	\nabla_{\bl} J^{\star} = \bmu^{\star}.
\end{equation}
\end{theorem}
The proof of this theorem is standard and can be found, for example, in~\cite{bertsimas1997lp}. The subgradient part of the statement comes from the fact that $J^{\star}$ is differentiable for almost all $\bl$, but not everywhere. If this is the case, $\bmu^{\star}$ is customarily taken as the (componentwise) largest vector in the set of subgradients.


\subsection{Solving DCOPF with known marginal prices}
\label{sec3-B}
The algorithm in~\cite{chen2020learning} uses a standard neural network to learn the cost function $J^{\star}$. Then $\bmu$ (the gradient with respect to the input) can be easily obtained through back propagation. For a nondegenerative LP problem, there would be exactly the same number of constraints as there are variables. Therefore, after using $\bmu^{\star}$ to find the active constraint sets, a linear system of equations is solved to find the optimal solution. 


We summarize the algorithm to solving problem (\ref{P1}) after the value of $\bmu$ is known in \textit{Algorithm 1}. This algorithm can offer an order of magnitude speedup compared to iterative solvers~\cite{chen2020learning}. 

\begin{table}[ht]
\normalsize
\centering
\begin{tabular}{ll}
\hline
\multicolumn{2}{l}{\textbf{Algorithm 1:  Solving DCOPF with given LMPs}}\\
\hline
\multicolumn{2}{l}{\textbf{Inputs:}~$\bmu$}\\
\textbf{Parameters:} $\bA$, $\bK$, $\bc$, $\bar{\f}$, $\bar{\bx}$\\
1:~Identify active nodal constraints using (\ref{gen-active})\\
2:~Identify active flow constraints using (\ref{flow-active})\\
3:~\textsf{EquationSolver}($\bx + \bA\f = \bl$, active constraints)\\
\textbf{Outputs:}~Optimal solutions $\bd{x}^{\star}$ to (\ref{P1})\\
\hline
\end{tabular}
\caption{Solving DCOPF for given LMPs.}
\label{algo:1}
\end{table}


\section{Network architecture and training loss design} \label{sec4}

The previous algorithm essentially states that if we can learn $\bmu^{\star}$ well, then we can obtain the optimal solution to (\ref{P1}). Note that $\bmu^{\star}$ need not be learned perfectly. Take the middle segment in Fig.~\ref{fig:single_bus} as an example. As long as the learned $J(l)$ has a derivative between 2 and 3 in this segment, we would detect the correct binding constraints. Therefore the key to success is to ensure that the learned  $\bmu^{\star}$ \emph{always} have small error.  

However, small training error does not guarantee small generalization error. The effectiveness of most machine learning algorithms are demonstrated through simulations, but engineering applications usually require some a priori guarantees, since well-trained models can fail to make reasonable inference on unseen input samples. 
In this section, we introduce two design features to guarantee the generalization ability of the trained model. By generalization, we mean a neural network that has a small training error in learning $\bmu^{\star}$ should have small testing errors on new samples. To guarantee generalization, we first utilize the convexity of the cost function to train an Input Convex Neural Network (ICNN). Then we leverage the Karush-Kuhn-Tucker (KKT) optimality conditions to design the training loss.


\subsection{ICNN architecture}
\begin{figure}[t]
\centering
\includegraphics[scale=0.4]{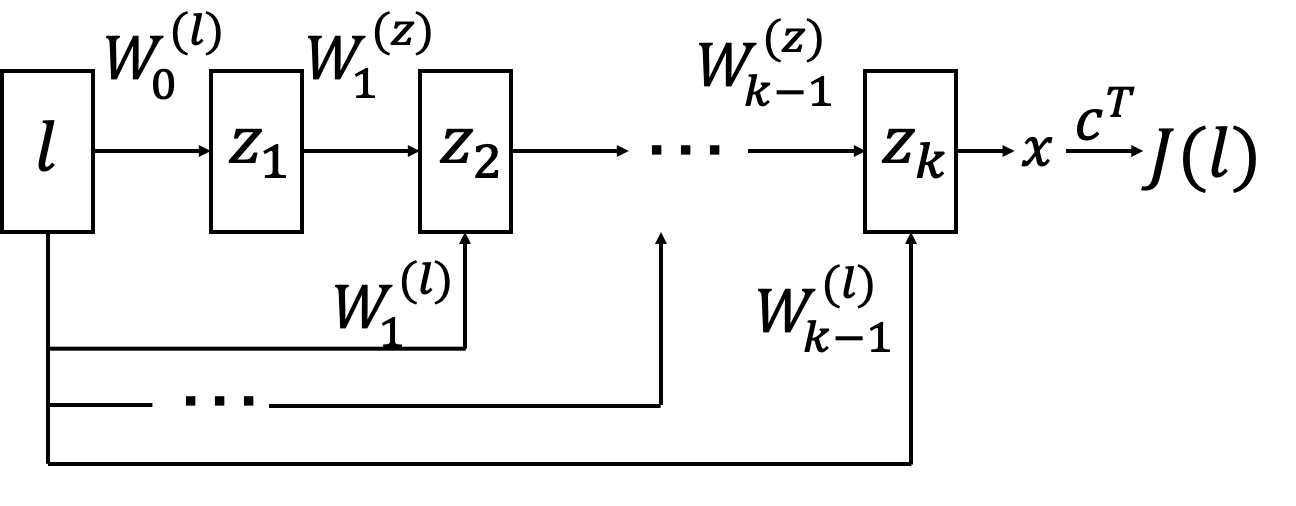}
\caption{The architecture of the trained ICNN. The weights $W_1^{(z)}, \dots, W_{k-1}^{(z)}$ are restricted to be nonegative. The pass through links $W_1^{(\bl)},\dots,W_{k-1}^{(\bl)}$ are not sign restricted.}
\label{fig:4-1}
\end{figure}

A useful result from linear programming that constrain the structure of $J^{\star}$ is that it is a convex function: 
\begin{theorem}
\label{convexity}
The optimal cost $J^{\star}(\bl)$ is a convex function of $\bl$ with its domain as the set of all feasible loads. .
\end{theorem}
The proof of this theorem is again standard and can be found in~\cite{bertsimas1997lp}. 
We adopt a special category of deep neural networks (DNNs), called Input Convex Neural Network (ICNN) to leverage this result~\cite{Chen19,Amos17}. The network architecture that we use in this paper is shown in Fig. \ref{fig:4-1}. The basic construction of ICNNs comes from composition of convex functions. Given two functions $f$ and $h$, if $f$ is convex and h is convex and nondecreasing, $g=h \circ f$ is convex. ICNNs satisfy this property by using ReLU as the activation functions and restricting the weights of the network to be nonnegative. This construction is shown to approximate all Lipschitz convex functions arbitrarily closely~\cite{Chen19}. 


\captionsetup[figure]{font=small,skip=2pt}

Suppose the fully-connected ICNN has $k+1$ layers, i.e., $k$ hidden layers with ``passthrough" and one extra linear output layer. The architecture can be written as follows:
\begin{align}
	\bz_{i+1} & = \sigma(\bW^{(z)}_i\bz_i + \bW^{(\ell)}_i\bl + \bb_i),~\textrm{for}~i=0, \cdots, k-1\label{icnn-1}\\
	\hat{J} & = \bc^T(\bz_k)\label{icnn-2},
\end{align}
where $\bz_{i}$ represents the output of the $i$-th hidden layer, $\bW^{(z)}_i$ represents the weight that connects the $i-1$-th hidden layer to the $i$-th hidden layer, and they are restricted to be nonnegative. The weights $\bW^{(\ell)}_{i-1}$ represent the matrices that directly connects the input $\bl$ to the $i$-th hidden layer. The symbol $\sigma(\cdot)$ represents the ReLU activation function. We interpret the result of the second to last layer to represent the generation solutions and multiply them by the cost vector $\bd c$ to get the total cost. 

We let $\btheta$ denote all trainable parameters in (\ref{icnn-1}) and (\ref{icnn-2}), i.e., $\btheta=\{\bW^{(z)}_{1:k-1}, \bW^{(\ell)}_{0:k-1}, \bb_{0:k-1}\}$, and the parameterized function $g_{\btheta}(\cdot)$ denote the mapping from $\bl$ to $\hat{J}$. Then (\ref{icnn-1}) and (\ref{icnn-2}) can be written in a more compact form
\begin{equation}
	\hat{J} = g_{\btheta}(\bl).
\end{equation}
The estimated value of $\bmu$ can be obtained by taking the derivative of $g_{\btheta}(\bl)$ with respect to $\bl$, and is denoted by $\hat{\bmu} = \nabla_{\bl}g_{\btheta}(\bl)$. We show in Section~\ref{sec5} that convexity is fundamental to the generalization property of the neural network. 

The goal of training the network is to learn the value of $\btheta$ which minimizes a specified loss function $\mathcal{L}$, i.e., 
\begin{equation}\label{icnn-task}
	\hat{\btheta} = \arg\min_{\btheta} \mathcal{L}\big(g_{\btheta}(\bl), \nabla_{\bl}g_{\btheta}(\bl)\big).
\end{equation}
Next, we illustrate how to construct this loss function. 

\subsection{Capturing KKT conditions}
When the load vector $\bl$ changes by certain amounts, the active constraint set and therefore the value of $\bmu^{\star}$ remain unchanged. In fact, we can divide the feasible input space of $\bl$ into a finite number of regions. Each of these region is a convex polytope and corresponds to a different combination of active constraints and a value of $\bmu^{\star}$. 

If we have a number of ground-truth values of $({J^{\star}}, {\bmu^{\star}})$ for every possible region, then we can train the neural network by minimizing the regression loss. However, the number of possible regions grows exponentially with respect to size of the system. For even a 5-bus system, it is unlikely that historical observations would include data in every region. Even if data is generated offline, exhaustively covering all of the regions with labeled data become cumbersome.  

If the data set only samples a small number of regions, the model trained by minimizing the regression loss performs poorly for input samples that come from unseen regions. This is not unexpected since there is no data to make prediction for these unseen regions. But in practice the value of using machine learning is to quickly determine what might happen for a large number of loads where some would come from unseen regions. Interestingly, because we are solving a well-defined optimization problem, we can mitigate this data challenge again by looking at duality theory. 

We develop an augmented training approach for the neural network $g_{\btheta}(\bl)$ based on violations of KKT conditions. Recall KKT conditions for the LP in (\ref{P1}) is
\begin{subequations} \label{eqn:kkt}
\begin{align}
&\bc - \underline{\btau} + \bar{\btau}-\bmu = \mathbf{0} \label{kkt-a}\\
&- \bK^T\underline{\blambd} + \bK^T\bar{\blambd} - \bA^T\bmu = \mathbf{0}\label{kkt-b}\\
&\mathbf{0} \leq \bx \leq \bar{\bx}\label{kkt-c}\\
&-\bar{\f} \leq \bK\f \leq \bar{\f}\label{kkt-d}\\
&\bx + \bA\f = \bl\label{kkt-e}\\
&\underline{\btau}, \bar{\btau}, \underline{\blambd}, \bar{\blambd} \geq \mathbf{0}\label{kkt-f}\\
&\underline{\tau}_ix_i = 0,~\bar{\tau}_i(x_i-\bar{x}_i) = 0,~\forall i\in\{1, \cdots, n\} \label{kkt-g}\\
&\underline{\lambda}_j(\bar{f}_j+K_j\f) = 0 \\
&\bar{\lambda}_j(K_j^T\f-\bar{f}_j) = 0,~\forall j\in\{1, \cdots, m\},\label{kkt-h}
\end{align}
\end{subequations}
where $K_j$ is the $j-$th column of matrix $\bK^T$.

Before introducing the loss term related to violations of KKT conditions, we have the following lemma:
\begin{theorem}
    The dual variables $\underline{\btau}, \bar{\btau}, \underline{\blambd}, \bar{\blambd}, \bmu$ satisfy the KKT conditions in \eqref{eqn:kkt} if and only if they satisfy equations \eqref{eqn:v14}:
    	\begin{subequations} \label{eqn:v14}
	\begin{align}
		[\bar{\blambd}+(\bK\f-\bar{\f})]^{+} - \bar{\blambd} = \mathbf{0}\label{v1}\\
		[\underline{\blambd}-(\bK\f+\bar{\f})]^{+} - \underline{\blambd}=\mathbf{0}\label{v2}\\
		[\bar{\btau}+(\bx-\bar{\bx})]^{+} - \bar{\btau}=\mathbf{0}\label{v3}\\
		[\underline{\btau}-\bx]^{+} - \underline{\btau}=\mathbf{0}\label{v4}.
	\end{align}
	\end{subequations}
\end{theorem}
\begin{proof}
Here we consider \eqref{v1} and the rest follow in similar fashions. In particular, we prove the following two conditions are equivalent: 
\begin{enumerate}
    \item $\bK\f\leq\bar{\f}, \bar{\blambd}\geq 0, (\bK\f-\bar{\f})\odot\bar{\blambd}=0$
    \item $[\bar{\blambd}+(\bK\f-\bar{\f})]^{+} - \bar{\blambd} = \mathbf{0}$
\end{enumerate}
where $\odot$ represents element-wise multiplication. 

First we show that 1) implies 2). If $\bK\f<\bar{\f}$, then we must have $\bar{\blambd}=0$ from the complementary slackness condition $(\bK\f-\bar{\f})\odot\bar{\blambd}=0$. So
\begin{equation*}
    [\bar{\blambd}+(\bK\f-\bar{\f})]^{+} - \bar{\blambd} = [\bK\f-\bar{\f}]^{+}=0.
\end{equation*}
If $\bK\f=\bar{\f}$ and $\bar{\blambd}\geq 0$, then we have
\begin{equation*}
    [\bar{\blambd}+(\bK\f-\bar{\f})]^{+} - \bar{\blambd} = [\bar{\blambd}]^{+}-\bar{\blambd}=0.
\end{equation*}

Next we show 2) implies 1). The right-hand-side implies the following dual feasibility since $[\bar{\blambd}+(\bK\f-\bar{\f})]^{+} - \bar{\blambd} = \mathbf{0}$ gives 
\begin{align*}
    \bar{\blambd}=[\bar{\blambd}+(\bK\f-\bar{\f})]^{+} \geq \mathbf{0}. 
\end{align*}
Suppose the primal feasibility does not hold, i.e., $\bK\f-\bar{\f}\geq 0$, then
\begin{align}
    &[\bar{\blambd}+(\bK\f-\bar{\f})]^{+} - \bar{\blambd} \\
    =&
    \bar{\blambd}+(\bK\f-\bar{\f})-\bar{\blambd}\\
    =&
    \bK\f-\bar{\f}\\
    =&0.
\end{align}
Therefore, we at most have $\bK\f=\bar{\f}$ and $\bK\f$ cannot exceeds $\bar{\f}$.


For complementary slackness,  suppose $\bK\f<\bar{\f}$ but $\bar{\blambd}\neq 0$, then we have
\begin{align}
    &[\bar{\blambd}+(\bK\f-\bar{\f})]^{+} - \bar{\blambd}\\
    =& 
    \left\{
\begin{array}{l}
      \either~ \bar{\blambd}+(\bK\f-\bar{\f})-\bar{\blambd}=\bK\f-\bar{\f}=0\\
      \tor~0-\bar{\blambd}=0,\\
\end{array} 
\right.
\end{align}
which contradicts our assumption. So, if $\bK\f<\bar{\f}$, we must have $\bar{\blambd}=0$. Suppose $\bar{\blambd}>0$ but $\bK\f<\bar{\f}$, we arrive a similar contradiction.
\end{proof}

Note that dual variables $\underline{\btau}, \bar{\btau}, \underline{\blambd}$, and $\bar{\blambd}$ can all be represented in terms of $\bmu$. To be specific,
based on (\ref{kkt-a}) and (\ref{kkt-f}), we can express the dual solutions $\underline{\btau}$ and $\bar{\btau}$ as follows
	\begin{subequations}
	\begin{align}
		\underline{\btau}(\bmu) & = [\bmu-\bc]^{+} - (\bmu-\bc)\\
		\bar{\btau}(\bmu) & = [\bmu-\bc]^{+},
	\end{align}
	\end{subequations}
	where the notation $[a]^{+}$ represents $[a]^{+}=\max\{a, 0\}$.
	The dual solutions $\underline{\blambd}$ and $\bar{\blambd}$ can be expressed in terms of $\bmu$ by solving the $\ell 1$-minimization problem in (\ref{P3}), and we denote them by $\underline{\blambd}(\bmu)$ and $\bar{\blambd}(\bmu)$, respectively.

Plugging the expressions $\underline{\btau}(\bmu)$, $\bar{\btau}(\bmu)$, $\underline{\blambd}(\bmu)$ and $\bar{\blambd}(\bmu)$ into (\ref{v1})-(\ref{v4}), we obtain a system of equations only related to $\bmu$.
To capture the KKT optimality conditions in (\ref{eqn:kkt}), we define \textit{violation degrees} associated with equations (\ref{eqn:v14}) as follows
\begin{subequations}
	\begin{align}
		\nu_{\bar{\lambda}} = [\bar{\blambd}+(\bK\f-\bar{\f})]^{+} - \bar{\blambd}\label{vd1}\\
		\nu_{\underline{\lambda}} = [\underline{\blambd}-(\bK\f+\bar{\f})]^{+} - \underline{\blambd}\label{vd2}\\
		\nu_{\bar{\tau}} = [\bar{\btau}+(\bx-\bar{\bx})]^{+} - \bar{\btau}\label{vd3}\\
		\nu_{\underline{\tau}} = [\underline{\btau}-\bx]^{+} - \underline{\btau}\label{vd4}\\
		\nu_{p} = \bl - \bx - \bA\f\label{vd5}.
	\end{align}
\end{subequations}
Based on \textit{violation degrees} given in (\ref{v1})-(\ref{v4}), we define the loss term to minimize violations of KKT conditions as follows
\begin{multline}\label{L_K}
\mathcal{L}_{\mathcal{K}}(\theta) = 
\|\nu_{\bar{\lambda}}(\grad)\|_2^2 + \|\nu_{\bar{\lambda}}(\grad)\|_2^2  \\
+ \|\nu_{\bar{\tau}}(\grad)\|_2^2
+ \|\nu_{\underline{\tau}}(\grad)\|_2^2 + \|\nu_{p}\|_2^2.
\end{multline}
We also define the regression loss as follows
\begin{equation}\label{L_R}
    \mathcal{L}_{\mathcal{R}}=\|g_{\btheta}(\bl)-J^{\star}\|+\|\grad-\bmu^{\star}\|,
\end{equation}
where the first term is the regression loss defined between $g_{\btheta}(\bl)$ and $J^{\star}$ over the neural network's outputs, and the second term is the regression loss between $\grad$ and $\bmu^{\star}$ over calculated derivatives.

By combining the the regression loss defined in (\ref{L_R}) and the KKT-related loss term (\ref{L_K}), we can express the training loss for our proposed algorithm as follows
\begin{equation}\label{L}
	\mathcal{L}(\theta) = \mathcal{L}_{\mathcal{R}}(\theta)+ \mathcal{L}_{\mathcal{K}}(\theta).
\end{equation}
By minimizing the loss function in (\ref{L}), we can get the trained model $g_{\btheta}(\bl)$ and use it to make predictions in an on-line way. We summarize our proposed algorithm for training the ICNN in Table \ref{table:4-1}, and call it \textit{Algorithm 2}.

\begin{table}[ht]
\normalsize
\centering
\begin{tabular}{ll}
\hline
\multicolumn{2}{l}{\textbf{Algorithm 2:~Algorithm for Training the ICNN}}\\
\hline
\multicolumn{2}{l}{\textbf{Inputs:}~Samples with labels $\mathcal{D}_{w}$}\\
\multicolumn{2}{l}{\textbf{Inputs:}~ samples without labels $\mathcal{D}_{w/t}$} \\
\multicolumn{2}{l}{\textbf{Inputs:}~Model to be trained $g_{\btheta}(\bl)$} \\
\textbf{Parameters:} $\bA$, $\bK$, $\bar{\f}$, $\bar{\bx}$, $\underline{\bx}$\\
1:~Calculate regression loss term $\mathcal{L}_R$ using (\ref{L_R})\\
\multicolumn{2}{l}{2:~Calculate KKT conditions-related loss term $\mathcal{L}_K$ using (\ref{L_K})}\\
3:~Minimize $\mathcal{L}(\theta)$ in (\ref{L}) to get the optimal parameter $\hat{\btheta}$\\
\textbf{Outputs:}~Trained model $g_{\hat{\btheta}}(\bl)$\\
\hline
\end{tabular}
\caption{Algorithm for training the ICNN.}
\label{table:4-1}
\end{table}




\section{Generalization} \label{sec5}

In this section we consider the generalization performance of our proposed method. By generalization, we mean the algorithm should perform well on test samples that were not seen during the training process. We adopt the standard method of analysis here: we assume that the neural network can be trained to zero error on the training samples, then we study the errors for testing samples~\cite{pineda1987generalization,andoni2014learning,Zhu2019}. We use the following definition as a shorthand:
\begin{definition}
We say a neural network is well-trained if it achieves zero loss on the training data. 
\end{definition}

\begin{figure}[ht]
    \centering
    \includegraphics[scale=0.5]{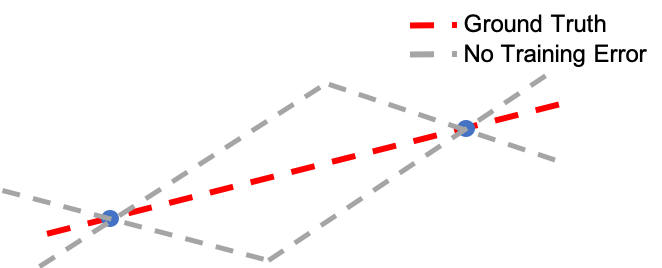}
    \caption{Given two points on a line, there are infinitely many piecewise linear functions passing through them. Therefore, small training error (passing through the points) does not necessary imply small generalization error (recovering the line).}
    \label{fig:example_generalization}
\end{figure}

Understanding the generalization properties is important because zero training error does not imply small test error. Consider the example given in Fig.~\ref{fig:example_generalization}. Suppose we are 
fitting a piece-wise linear function but only given
two points that lie on a line and the training loss is the regression loss. There are infinitely many functions that pass through these points, implying that they have zero training error. Obviously, many of them can have large testing errors for other points on the line. This example also shows that it is not sufficient to just impose convexity or provide gradient information on their own. There are also infinitely many convex functions passing through the labeled points, and there are infinitely many functions with the right gradients at the given points. To constrain the class of functions to be learned, both convexity and the gradient information are needed. 

Since $J^{\star}$ is piece-wise linear, we study generalization for two settings. The first is we assume that there are multiple training data within a region where $J^{\star}$ is linear, and we are interested in the testing performance of a new input from the same region. This setting is about whether the model is constrained enough to not overfit during training. We provide a positive (and simple) answer in Theorem~\ref{thm:same_region}. 

The second setting is to assume that the test data lies in a region that was unseen during the training process. 
This question is normally not asked since there is no expectation that learning would be useful for these type of unseen data. However, since there are exponential number of LP regions for DCOPF, it is likely that not all regions would be included in the training data. Therefore, learning algorithms must provide some guarantees on unseen regions. In Theorem~\ref{thm:unseen_region}, we provide a positive answer showing that the gradients of the unseen region are still bounded. 


\subsection{Generalization for A Linear Region}
\label{sec5A}
In this part, we study the case where all training samples have the same value for $\bmu$. That is, they come from the same LP region. Let us denote the set of training samples as $\mathcal{D}_{trn}$, and $\conv \mathcal{D}_{trn}$ is the convex hull of set $\mathcal{D}_{trn}$. The following theorem states the performance of the neural network for a new input $\bl^{\textrm{new}}$ that is not in $\mathcal{D}_{trn}$.

\begin{theorem}\label{thm:same_region}
Given $N$ input loads $\mathcal{D}_{trn}=\{\bl^1,\dots,\bl^N\}$ and assume $\hgradi = \bmu$ for all $i=1,\dots,N$. Assume the ICNN model $g_{{\btheta}}(\bl)$ is well-trained on $\mathcal{D}_{trn}$. Then for all points $\bl^{\textrm{new}} \in \conv \mathcal{D}_{trn}$,  $\hgrad = \bmu$.
\end{theorem}
This theorem is useful for LP problems because the gradient of $J$ is piecewise constant over convex polytopic regions. If we are given some points from a region, then a well-trained neural network guarantees that the function is learned correctly for all points within their convex hull. This result implies the correctness of the overall algorithms since they only rely on the gradient (dual variable) information.

Technically, the above theorem says that if the gradients of a convex function are equal at a set of points, then the function is linear on the convex hull of these points. This is not a surprising result, but it does show that by constraining the structure of the neural network and using gradient information, we can generalize to uncountable number of points (compact regions) by learning from a finite number of training points. The proof of the theorem is given in Appendix~\ref{app:same_region}. 

\subsection{Generalization with KKT Loss}
If training samples may not represent all possible regions, we can construct the KKT conditions augmented loss in (\ref{L_K}) on a set of unlabeled data points. Since we do not ask for labels, 
we can sample as many data points as we want. This allows us to train with a very large set.
We call this set of unlabeled data points the helper set.

By training with the helper set, unseen regions become ``seen" in the sense that the outputs from the trained model must satisfy KKT conditions. As long as the model is well-trained, the analysis of generalization is the same as Section \ref{sec5A}. Therefore, the generalization performance of training with helper set can also be guaranteed by Theorem \ref{thm:same_region}.

\subsection{Generalization for Unseen Regions}
Here we consider the case where a region is not represented at all by the training data, including both labeled and helper sets. Then if a test sample comes from this region, would our method output anything useful? Methods like classification and dictionary learning cannot make useful predictions since there is no basis to make inferences about unseen regions. The next theorem shows that our approach is still partially successful because the gradient of a test data point is bounded by the gradient of the training points:
\begin{theorem}\label{thm:unseen_region}
Given $N$ input loads $\mathcal{D}_{trn}=\{\bl^1,\dots,\bl^N\}$, assume the ICNN model $g_{{\btheta}}(\bl)$ is well-trained on $\mathcal{D}_{trn}$. Assume that $N\geq n+1$ and $\mathcal{D}_{trn}$ does not lie in a lower dimensional subspace in $\R^n$. Then for all points $\bl^{\textrm{new}} \in \conv \mathcal{D}_{trn}$,  $\hgrad$ is contained in a bounded convex polytope. 
\end{theorem}
The exact characterization of the polytope is given in the proof in Appendix~\ref{app:unseen} and depends on the values of $\{\bl^1,\dots,\bl^N\}$ and $\hgrad$. The significance of the theorem lies in that training data are able to constrain the gradient for all points that lies in its convex hull. Intuitively speaking, as long as some surrounding points are included in training, the gradient cannot be ``very wrong" even for points coming from LP regions that were not seen during training. 

\begin{figure}[ht]
    \centering
    \includegraphics[scale=0.4]{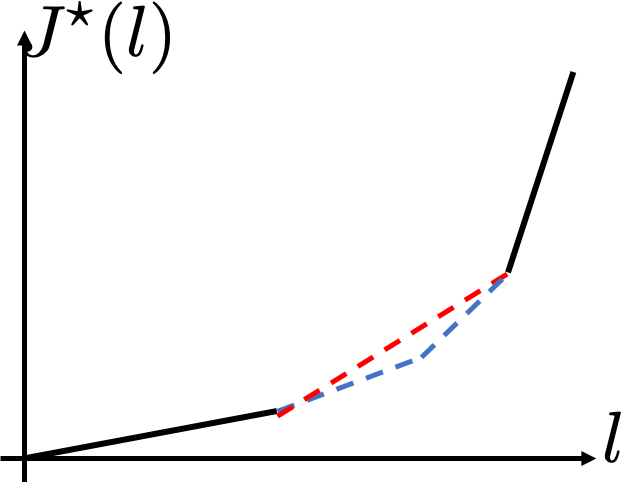}
    \caption{Example when the middle region has no data. But as long as the other two regions are well trained (black lines), the slopes in the middle region are bounded (blue and red dashed lines) by Theorem~\ref{thm:unseen_region}. Then the active constraint detection (Algorithm 1) would still be correct.}
    \label{fig:unseen}
\end{figure}

Consider the curve in Fig~\ref{fig:single_bus}. Suppose we learned the two end pieces correctly but there was no training data for the middle piece, as shown in Fig.~\ref{fig:unseen}. Then by Theorem~\ref{thm:unseen_region}, the slope of the middle piece is constrained to be between the slope of the two end pieces. Furthermore, even if the neural network is trained in such a way that there are more than one piece of the middle region, the slopes of all of the pieces are still bounded between the two end pieces. Since Algorithm 1 only relies on getting $\mu$ to be in the correct range, the active constraints would be identified correctly for all of these cases. 

Theorem~\ref{thm:unseen_region} formalizes the picture in Fig.~\ref{fig:unseen} to higher dimensions, but the geometric intuition remains the same.   This theorem also formalizes the empirical observation in~\cite{venzke2020learning}, where the error of neural network-based OPF is reduced if training points are on the boundary of the feasible region.

\section{Experimental results} \label{sec6}
\begin{table*}[ht]
\normalsize
\centering
\begin{tabular}{ c c c c || c c c  }
\hline
Input variations & \multicolumn{3}{c||}{$30\%$} & \multicolumn{3}{c}{$50\%$} \\
\hline
Solutions & Optimality & Feasibility & Infeasibility & Optimality & Feasibility & Infeasibility  \\
\hline
End-to-End & $18.34$ & $21.98$ & $78.02$ & $17.76$ & $46.30$ & $53.70$ \\
\textbf{Our model} & $\textbf{94.93}$ & $\textbf{94.93}$ & $\textbf{5.07}$ & $\textbf{93.08}$ & $\textbf{93.08}$ & $\textbf{6.92}$ \\
\hline
\end{tabular}
\caption{Quality of solutions. We compare the solutions quality of our model to the end-to-end model. 
The ratios of optimal, feasible and infeasible solutions are listed and the numbers represent percentages. Results under two different input variations are given, i.e., $30\%$ and $50\%$ deviations from the nomial load value. More than $90\%$ of the solutions obtained from our algorithm are optimal, while less than half of the solutions from the end-to-end model are feasible.}
\label{table:6-1}
\end{table*}
In this section, we demonstrate experimental results of using Algorithm 2 for training the ICNN and Algorithm 1 for solving DCOPF. We use the IEEE 14-bus system as the benchmark. We first examine the quality of solutions in terms of feasibility and optimality, then we examine the generalization performances.

To generate the training set, we first sample $\bl$ from the uniform distribution. We use two different variations from nominal load values: $30\%$ and $50\%$. The total size of data samples is $50000$ for each setting. Then, for each value of $\bl$, we solve the primal and dual problems using CVXPY \cite{diamond2016cvxpy} powered by CVXOPT \cite{cvxopt}. The optimal cost and the dual solutions are recorded. We hold $20\%$ of all data samples as the test set and use the remaining for training. The ICNN we train has $4$ hidden layers.

In order to better evaluate the performance of the model trained using our algorithm, we also provide experimental results of the end-to-end learning model, which is commonly used in the field of solving DCOPF \cite{pan2019deepopfb}. In the end-to-end model, we train a 4-layers fully-connected ReLU network by minimizing the regression loss, and use the trained model to directly predict the optimal solution to (\ref{P1}). We keep training both models until the training loss converges. In comparison to the end-to-end model, it takes longer time to train our model for each epoch. This is because we use Tensorflow package to create the end-to-end model, while we have to implement our model manually.

\vspace{-0.5cm}
\subsection{Overall performance}
To evaluate the quality of solutions obtained by different learning models, we divide solutions into three categories:~optimal, feasible, and infeasible solutions. In our model, we feed $\bl$ into the neural network and find the active constraints set. Then we solve a system of linear equations using a standard solver to obtain final solutions. The end-to-end model can directly outputs the solution to $\bx$.
 The ratios of optimal, feasible and infeasible solutions obained by different learning models are listed in Table \ref{table:6-1}. As shown in Table \ref{table:6-1},
the optimality ratio of the solutions obtained from our model is higher than $90\%$ under both input variations. In comparison, almost $50\%$ of the solutions obtained from the end-to-end model are infeasible. In terms of computational time, both methods are much faster than iterative solvers in online DCOPF solving.

To examine the solution feasibility in the end-to-end model, we use the output $\bx$ and the nodal power balance (\ref{1c}) to obtain $\f$. The value of feasibility ratio depends on how large the error tolerance is. In this paper, $0.3\%$ mismatch is allowed when we calculate feasibility ratios. 
In Table \ref{table:6-3}, we also list the ratios of solutions that do not satisfy the nodal power balance, the limits on generators' outputs, and the limits on line flows, respectively. From Table \ref{table:6-3}, we can see that more than $98\%$ of the solutions obtained from our model satisfy both generators limits and line flows limits.
By contrast, only $24\%$ of the solutions from the end-to-end model satisfy the line flow limits.

\begin{table*}[ht]
\normalsize
\centering
\begin{tabular}{ c c c c || c c c  }
\hline
Input variations & \multicolumn{3}{c||}{$30\%$} & \multicolumn{3}{c}{$50\%$} \\
\hline
Infeasibility & Nodal balance & Generators limits & Lines limits & Nodal balance & Generators limits & Lines limits  \\
\hline
End-to-End & $16.88$ & $0$ & $76.79$ & $47.45$ & $6.33$ & $6.46$ \\
\textbf{Our model} & $\textbf{5.07}$ & $\textbf{0}$ & $\textbf{1.18}$ & $\textbf{6.92}$ & $\textbf{0}$ & $\textbf{0.79}$ \\
\hline
\end{tabular}
\caption{Infeasibility of solutions. We compare our model to the end-to-end model.
The ratios of solutions that do not satisfy the nodal power balance, the limits on generators' outputs, and the limits on line flows are listed, respectively. $0.3\%$ mismatch is allowed. We give results under two different input variations, i.e., $30\%$ and $50\%$. In both settings, more than $98\%$ of the solutions obtained from our model satisfy both generators limits and line flows limits, and more than $90\%$ of our solutions satisfy the nodal power balance.}
\label{table:6-3}
\end{table*}

\subsection{Generalization on Unseen Regions}
\begin{figure}[t]
\centering
\includegraphics[scale=0.35]{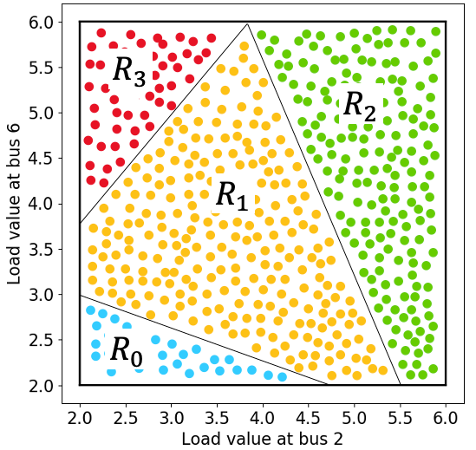}
\caption{Division of the input space. The axes are load values at the two buses. In this example, based on different combinations of active constraints, the input load space can be divided into four regions. We take samples from $R_1$ as the testing set and samples from surrounding regions $R_0$, $R_2$ and $R_3$ as the training set.}
\label{fig:6-1}
\end{figure}

To evaluate the generalization ability of our model, we create an illustrative example in the 14-bus system. Particularly, we examine the generalization performance on new data points that comes from region without any training samples. To generate the training set for this case, we only change the load values at two buses, but keep the remaining load values fixed. In this way, the space of input loads can be regarded as a two-dimensional plane. When varying the load values at the two buses, we can have four different combinations of active constraints, which correspond to four different values of $\bmu^{\star}$. Therefore, we divide the input load space as four regions, denoted as $R_0$, $R_1$, $R_2$ and $R_3$. The division of the input space is shown in Fig. \ref{fig:6-1}. We take training samples from $R_0$, $R_2$, and $R_3$, and take testing samples from $R_1$. Let us denote the training set as $\mathcal{D}_{trn}$, and the testing set as $\mathcal{D}_{tst}$.

\captionsetup[figure]{font=small,skip=2pt}

We use two different training approaches for our model. In the first training approach, we only use $\mathcal{D}_{trn}$ for training and minimize both the regression loss and the KKT-related loss on $\mathcal{D}_{trn}$. For the second training approach, we construct an additional training set, called helper set $\mathcal{D}_{help}$.
To generate the helper set, we can sample $\bl$ from uniform distributions $\ell_2\sim$~\textsf{Uniform}$(2, 6)$ and $\ell_6\sim$~\textsf{Uniform}$(2, 6)$. Therefore, $\mathcal{D}_{help}$ contains the testing region $R_1$.
Aside from minimizing (\ref{L}) on $\mathcal{D}_{trn}$, we also minimize the KKT-augmented loss (\ref{L_K}) on $\mathcal{D}_{help}$. Note that $\mathcal{D}_{help}$ is not labeled. End-to-end model only use labeled samples and is trained on  $\mathcal{D}_{trn}$.

\begin{table}[ht]
\normalsize
\centering
\begin{tabular}{ c c c c }
\hline
& Optimality & Feasibility \\
\hline
End-to-end & 5.52 & 8.6 \\
\textbf{Our model, with} $\bd{\mathcal{D}_{help}}$& \textbf{97.24} & \textbf{97.24} \\
\textbf{Our model, without} $\bd{\mathcal{D}_{help}}$& \textbf{62.25} & \textbf{72.31} \\
\hline
\end{tabular}
\caption{Generalization performance on test samples coming from never seen regions.  With the helper set, our method is optimal $97\%$ of the time. Even without the helper set, the optimality ratio is 62\%. The end-to-end model fails to make reasonable predictions.}
\label{table:6-2}
\end{table}


We list the ratios of optimal, feasible and infeasible solutions obtained from different learning models in Table \ref{table:6-2}. As we can see, when we use $\mathcal{D}_{help}$ as an additional training set, we can obtain an optimality ratio as high as $96\%$. Even without $\mathcal{D}_{help}$, more than half of the solutions obtained from our model can achieve optimal values. 
As a comparison, the end-to-end model fails to make feasible predictions on test samples that come from never seen regions in the training process. The reason that our proposed algorithm outperforms the end-to-end model can be attributed to the KKT-related loss. By minimizing the KKT-related loss term, the trained model is able to learn the underlying KKT conditions in all four regions and make better predictions of $\bmu^{\star}$ on $\mathcal{D}_{tst}$.



\section{Conclusions}

This paper proposes a new framework to use neural networks for solving DCOPF. By leveraging rich linear programming theories, we prove our framework guarantees generalization.
First, using the convexity of optimal cost in DCOPF, we constrain the neural network to have an input convex structure. Second, using the KKT optimality conditions, we add violations of KKT conditions to the training loss.
In this way, we are able to exploit large amounts of unlabeled data points for training and improve the generalization performance. Our method is evaluated on the IEEE 14-bus data set and compared to end-to-end learning models. The experimental results demonstrate that our method can increase the optimality ratio of solutions by a factor of five, and decrease the infeasibility ratio by up to an order of magnitude.

\bibliographystyle{IEEEtran}
\bibliography{mybib.bib}

\vspace{-0.5cm}
\appendix
\subsection{Fundamental Flows} \label{app:flow}
In the DC power flow model the power flow on the lines are determined by the angle differences. Let $\theta_i$ be the angle of bus $i$. Let $f_{ij}=b_{ij}(\theta_i-\theta_j)$ be the flow along the line connecting $i$ and $j$. If a network has cycles, let buses $1,,\dots,n_c$ be the buses in a cycle, counted in either clockwise or  counterclockwise direction. Consider the weighted sum
\begin{align*}
    & \frac{f_{12}}{b_{12}}+\frac{f_{23}}{b_{23}}+\dots+\frac{f_{n_c 1}}{b_{n_c 1}} \\
    =&(\theta_1-\theta_2)+(\theta_2-\theta_3)+ \dots+(\theta_{n_c}-\theta_1) \\
    =& 0. 
\end{align*}
Therefore, the flows lie in a subspace. 

Repeating the above calculation for every cycle in a network gives that the flows lie in a subspace of dimension $n-1$ for a connected network with $n$ buses. A basis of this subspace is called a set of fundamental flows. There are multiple bases to choose the fundamental flows from. A popular way is to choose a spanning tree and consider the flows on the branches as fundamental, and everything else can be derived from them. 



\subsection{Proof of Theorem~\ref{thm:same_region}} \label{app:same_region}
\begin{proof}
Since $\bl^{\textrm{new}}$ is in the convex hull of $\mathcal{D}_{trn}$, there are positive coefficients $\alpha_1,\dots,\alpha_N$ such that 
\begin{equation*}
    \bl^{\textrm{new}} = \alpha_1\bl^1 +\cdots + \alpha_N\bl^N,
\end{equation*} 
and $\alpha_1+\dots,\alpha_n=1$. By convexity, 
\begin{equation}\label{samereg:leq}
    g_{{\btheta}}(\tell)\leq \alpha_1 g_{{\btheta}}(\bl^1) + \cdots +  \alpha_{N}g_{{\btheta}}(\bl^{N}).
\end{equation}
By the assumption that $g_{{\btheta}}(\bl)$ is well-trained, we have 
\begin{equation}
    \gradi=\bmu, \textrm{~for~} i=1, \cdots, N. 
\end{equation}
Using first-order conditions of convex functions, we have
\begin{subequations}
\begin{align*}
    g_{{\btheta}}(\tell) & \geq  g_{{\btheta}}(\bl^1) +  \bmu(\tell-\bl^1)\\
        & \dots\\
 g_{{\btheta}}(\tell) & \geq  g_{{\btheta}}(\bl^N) +  \bmu(\tell-\bl^N).\\
\end{align*}
\end{subequations}
Multiplying the $i$'th equation by $\alpha_i$ and summing gives
\begin{equation}\label{samereg:geq}
    g_{{\btheta}}(\tell) \geq \alpha_1 g_{{\btheta}}(\bl^1) + \cdots +  \alpha_{N}g_{{\btheta}}(\bl^{N}).
\end{equation}
Combining \eqref{samereg:leq} and \eqref{samereg:geq} gives 
\begin{equation}
    g_{{\btheta}}(\tell)= \alpha_1 g_{{\btheta}}(\bl^1) + \cdots +  \alpha_{N}g_{{\btheta}}(\bl^{N}).
\end{equation}
This implies the function is linear in the convex hull of $\mathcal{D}_{trn}$ and all the points have the same gradient. 
\end{proof}

\subsection{Proof of Theorem~\ref{thm:unseen_region}} \label{app:unseen}
Suppose $g_{\btheta}:\R^n \rightarrow \R$ is a convex function. Given $\bl^1,\dots,\bl^N$ in the domain of $g$ and let $\bmu^i=\gradi$. Let $\tell$ be a point in the convex hull of $\bl^1,\dots,\bl^N$ and denote $\hgrad=\bmu$. By the convexity of $g_{\btheta}$, we have 
\begin{equation} \label{eqn:bound}
(\gradi-\hgrad)^T (\tell-\bl^i) \leq 0, \forall i. 
\end{equation}
The inequalities in \eqref{eqn:bound} constrain the values that $\hgrad$ can take. We are to show that these inequalities actually describe a closed and bounded polytope in $\R^n$. We do this by contradiction.

Suppose the region defined by the inequalities in \eqref{eqn:bound} is not bounded. Then $\hgrad$ can be scaled arbitrarily and all of the inequalities in \eqref{eqn:bound} would still hold. Then we can take the norm of $\hgrad$ to be large enough such that it would dominate the $\gradi$ terms. Then \eqref{eqn:bound} becomes 
\begin{equation} \label{eqn:bound2}
\hgrad^T (\tell-\bl^i) \geq 0, \forall i. 
\end{equation}
Since $\tell$ is in the convex hull of $\bl^1,\dots,\bl^N$, we can write it as $\tell=\alpha_1 \bl^1+\dots+\alpha_N \bl^N$ and $\alpha_i \geq 0$ and sums up to 1. Substituting this into \eqref{eqn:bound2} and rearranging the terms, we have 
\begin{equation*}
    \alpha_1\hgrad^T \bl^1+ \dots+ \alpha_N\hgrad^T \bl^N \leq \hgrad^T \bl^i, \forall i. 
\end{equation*}
By the assumption that $N\geq n+1$ and $\bl^1,\dots,\bl^N$ are not in a lower dimensional subspace of $\R^n$, $\hgrad^T \bl^i$ will be nonzero for at least two $i$'s. But it is not possible to have a convex combination of scalars (the $\hgrad^T \bl^i$'s) larger than every scalar in the set when at least two are nonzero~(this follows from Farkas' lemma). Thus we have a contradiction from the assumption that the polytope created by \eqref{eqn:bound} is unbounded.

\end{document}